# SOME EPISTEMIC QUESTIONS OF COSMOLOGY

## PETAR V. GRUJIĆ

Institute of Physics, PO Box 57, 11080 Belgrade, Serbia
*Tel. 381-11-3162-099; fax: 381-11-3162-190; e-mail: grujic@phy.bg.ac.yu*

ABSTRACT. We discuss a number of fundamental aspects of modern cosmological concepts, from the phenomenological, observational, theoretical and epistemic points of view. We argue that the modern cosmology, despite a great advent, in particular in the observational sector, is yet to solve important problems, posed already by the classical times. In particular the stress is put on discerning the scientific features of modern cosmological paradigms from the more speculative ones, with the latter immersed in some aspects deeply into mythological world picture. We finally discuss the principal paradigms, which are present in the modern cosmological studies and evaluate their epistemic merits.

KEY WORDS: cosmology, epistemology, methodology, mythology, philosophy of science

## 1. PROEM

Cosmos appears a very large object of study indeed, by all standards – epistemic, observational, theoretical and phenomenological ones (see, e.g. Ellis, 1991, 1999). It may well be argued that it is too large object for rational examinations, as compared with its parts that are subject to scientific enquiries. In this context the very notion of Cosmos, in the strict sense, appears inadequate, for it presupposes something that is to be investigated – the concept of order, or system. In the following we shall use the term Universe, as a more general designation of the World in its totality. Further, many terms, definitions and concepts in general appear insufficiently general to deal with the subject, since they are associated, albeit implicitly, with particular cosmological models, even paradigms. Anthropic principles, genesis, eschatology, evolution, cosmic (arrow of) time etc are examples in point, since they appear senseless in a number of particular cosmological paradigms, as we shall see later on.

Aristotle already was aware that the whole is not just a sum of its parts. This aspect of epistemic and phenomenological studies poses serious problems in studying the Universe, an "object" which we shall never have an experience of. Many other concepts, like that of completeness, acquire particular significance when talking on the World as a whole. From the methodological pint of view, conceptual and phenomenological features appear so strongly entangled in cosmology, that a deductive approach is doomed to fail from the very start, as illustrated by the choice of concepts and cosmological paradigms above. All aspects mentioned above – phenomenological, theoretical, epistemic and observational ones, appear tightly interconnected, that even the analytical method, so dear to our inductive procedure in studying complex systems, fails in the last analysis.

All these features explain why our understanding of the Universe, despite the enormous progress at the technical (including mathematical) level, turns out not fundamentally different from the inferences of ancients, in particular antique Greeks. This, however, should not be conceived as our failure, but rather as our appreciation of the fundamental character of the cosmology as such.

## 2. INTRODUCTION

As a research field cosmology covers a broad range of physical conditions (ontological aspect) and wide spectra of methodological approaches (epistemological aspects). The former go from the visually accessible parts of the cosmos we live in, to the remote parts whose nature we may only speculate about. It is these inaccessible regions that force us to resort to (often wild) speculations that go beyond the positive science. These speculations, though dressed in mathematical clothes, have provoked a number of astrophysicists, including cosmologists, to compare modern cosmology with traditional mythological representations of the universe.

On the more positive side, our experience, both rational and intuitive stems from the empirical inference from limited compartments, as parts of the envisaged whole. An extrapolation to the entire physical reality meets insurmountable difficulties, not so much of a quantitative, but predominantly qualitative nature. The procedure adopted by cosmologists in widening the scope of the issue is of the trial and error kind. It is this approach that fills the "speculative phase space" of the cosmological studies. As the empirical inference extends and some of many speculative models are added to the standard paradigm (whatever it might mean), the science fiction part of the cosmological studies becomes positive science. Nevertheless, proliferation of the speculative, all-embracing models reminds us of the pre-Socratic *fisikoi* and their bold speculations on the ultimate nature of the world. That some of those speculations have provided us with a number of very valuable concepts, like the atomistic one, for instance, or the hierarchical cosmos, shows how the free play of human mind furnish the evolutional path of the mankind with powerful tools.

To make our exposition as clear as possible, we shall adopt the following terminology in dealing with a universe. If we refer to an ideal concept, or notion of the totality of a world, we use capital initials, like Universe, Cosmos etc. It will not necessarily be associated with our universe (the Universe), not even with our model of it, but rather as a "divine" view, or a Platonic concept.[1] Hence, by a Universe we shall mean a totality of the physical reality, irrespective of the notion of order or disorder. Within a Space-Time manifold, it may comprise ordered parts (systems), which we call Cosmoses. Note that notion of multi-Universe (Multiverse) makes no sense within this vocabulary, nor Genesis, End, etc. In the similar vain, if we mention God it will not imply the God, that is any particular concept of this entity, especially not any particular historical, theological and similar specification, certainly not under a definite name, even not an abstract god like that due to Espinoza, for example. Further, by Mythology we shall refer to a mythology as a language, as defined by Barthes, for example (Barthes, 1957), and not to any specific one, like Greek (Homeric, Hesiodic), Judaistic (Biblical, Cabbalistic) etc.

In exposing various models and paradigms, we shall not follow chronological order, for this would imply a resort to the notion of an evolution, which is pertinent to some of the classes of cosmological models, but not to all extant. The very concept of Evolution appears one of the principles, albeit in a disguised form, in Cosmology, which should be treated on equal footing with other cosmological principles, to be considered later on.

In the next section a number of epistemological questions are discussed as a prerequisite to their more detailed account with regard to the actual cosmology. A brief exposition of the principal theoretical background (mainly General Relativity and Quantum Mechanics) of the modern cosmology is given in section 4, with an emphasis on the question of the interrelation between their theoretical structure and domain of their applicability. Section 5 is devoted to relations between different cosmic strata and the relevant epistemological and inspirational issues. The question of uniqueness, as a specific (and exclusive) cosmological issue is addressed in section 6. A number of actual and possible inspirations as sources for particular cosmological models are discussed in section 7, as well as the epistemological values of nonscientific associations. In section 8 we enumerate and discuss principal classes of cosmological models, with a brief historical account of their development. In the next, $9^{th}$ section the issue is further elaborated, in more general terms as well as applied to particular paradigms. The issue of the uniqueness is considered again in section 10, from a more practical perspective, discussing the relevant cosmological offers on the market. What are differences between positive and speculative science is elaborated in section 11, as illustrated by the current leading cosmological paradigm, that of Standard Model. The problem of empirical verification of an assumed cosmological reality is addressed in section 12, with an emphasis on the possible empirical illusions, as met in the observational cosmology. Finally, we summarize our findings and discussions in the concluding, last section.

## 3. COSMOLOGICAL EPISTEMOLOGY

If one restricts himself to the Western tradition, two principal periods of the cosmological studies can be distinguished: (i) Pre-Socratic (7-5 c. BC) and (ii) 20-ieth century. Almost all contemporary cosmological paradigms could be traced back to the pre-Classical Hellada, as we shall see immediately. But before doing that, it is convenient to start from a more basic notion of *cosmological principles,* as formulated explicitly within modern studies of the Universe.

### 3.1 Cosmological ontological principles

These serve several purposes. First, they help classify a plethora of particular models, according to some common features. Second, having essentially approximate nature, these principle help conceiving physical cosmic reality in a more transparent, albeit inaccurate description. Last, but not the least, they enable one to solve mathematical equations as central constructions of any modern cosmological model. We shall distinguish the following principles.
  (a) Homogeneity of space (uniform matter distribution)
  (b) Homogeneity of time (structure independent of the cosmic global time)
  (c) Isotropy of space (independence of structure on the direction of observation)
  (d) Homothety of space (independence of structure on scale transformations)

A particular cosmological paradigm comprises a number of the above principles. Thus, the Standard (Hot Big Bang) model includes (a,c), Steady State model (a,b,c), Hierarchical model (c,d), etc. What the universe has in common with these Paradigms (ontological aspects) and how one links the observational evidence with Models (epistemological aspects) will be our main concern in the following. But before doing that a few words on the nature of our inquiries are in order.

### 3.2. Physics, metaphysics and philosophy

Cosmology in its strict sense is physics and as such belongs to empirical sciences, often identified by "hard", "positive" etc science. Metaphysics implies a wider domain of inquiries, which extends beyond the empirical evidence, resorting often to speculative inference. It is here that the epistemological analysis helps sorting out credibility of offered explanations of the nature of empirically hardly accessible (or even inaccessible in principle) reality. Even further from the domain of positive evidence is the philosophical study, which provides a general framework for interpreting inferences that go beyond positive science, even science as such. It is this domain where notion like purpose, teleology (providence), eschatology etc enter the arena of human speculations. Of course, one may go one step further, and dive into theology, as "the highest philosophy" as Aristotle put it, but this would be beyond the scope of our interest here.

### 3.3. Cosmological epistemological principles

Our cosmos is very large physical system by all standards. The latter are formed, in their turn, after our experience within a limited portion of supposed larger universe. What defines the largeness of a system? It is the typical time period necessary for exchanging informations compared with typical lifetime of an observer (or of a limited number of generations, at most). Practically, it is the speed of transmitting informations, as exemplified by the speed of light $c$, which measures distances of astronomical scales. We are aware that by looking into remote cosmic regions we are looking into the past. Formally, we live in a small portion of the total space-time manifold, and any extensions beyond the observable cosmos are of necessity speculative ones. In the previous sections we mentioned cosmological principle, which should guide us in inferring inaccessible regions of the universe. But by carrying out our estimate as for the properties of unobserved, we make use of, albeit implicitly, another set of principles, we call cosmological epistemological principles. One may, thus make use of the same principles, as quoted above, but applied to our scientific, or ideological, background, on which we construct our ontological assumptions.

We assume, hence, that the physical laws, as we know them here on Earth and our planetary system, are valid to large, hitherto unexplored distances, both in space and time. More over, even in inferring astrophysical evidence of the accessible parts of the universe, we assume tacitly that the same physical laws govern those remote regions. In fact, the only principle, which need not be invoked in that, is that of the space isotropy. Incidentally, it is the only property we assume about cosmos that is easily verifiable (and is verified up to a large accuracy). Since our inference into the physical properties and phenomena is always indirect and linked inexorably with our theoretical models, the empirical evidence rests largely on the validity of these theoretical constructions. The case in point is estimate of distances of remote celestial objects, especially extragalactic ones. The primary observational quantity is the colour shift of spectral lines received from these objects (if they are luminous, of course). The rest relies on the additional interpretation of this shift (typically red shift). Whether this is attributed to Doppler (kinematic) effect, to gravitational (dynamic) phenomena, space dilation (geometric attribute) etc, depends our interpretation of the dynamical state of the universe. But even then we assume that the emitters (atoms, for instance), at the place and (cosmic) time when emitted, correspond to our models, which involve a number of universal, fundamental constants, like Planck's $h$, speed of light (and any other fundamental dynamic field) $c$, Newton's gravitational constant $G_N$, etc. The assumption that there exist fundamental, universal constants is not a choice, but postulate by necessity. As every postulate, they are Achilles' heels of our theoretical constructions, the alternative choices being cut by Okham's razor.

Hence, by exploring remote regions in space and time, we assume these regions are just like our small parts, which are accessible to the experimental testing. The rest, accessible to the observational evidence is supposed to obey Copernican principle, which maybe considered to be the generic form of the first cosmological principle (a), as mentioned above. The forth principle (d) stands somewhat apart from the rest. First, it is only recently, with advent of the so-called hierarchical paradigm that the principle came to light. Second, because of the essentially quantum nature of the microworld, small distance physics (or alternatively, high-energy physics), indicates, according to all field-theoretical model, the breakdown of the homothety principle at so-called Planck's scales. At these scales the very notion of space and time looses meaning and the space-time manifold becomes, presumably, deprived of any discernable structure, what makes scaling transformations meaningless too. Hence, within the "epistemic space", ontologically defined principles (a,b,c) are postulate, but the fourth one (d) surely breaks at sufficiently small scales. It is within these scales that one expects the unification of all fundamental interactions, including presumably the gravitational one.

## 3.4. Ontological prerequisites

Under above assumptions we expect to deal with at least a part of the observable cosmos, which can be treated in ordinary manner, as a familiar physical system. This part of cosmology, which may be dubbed cosmography, can be regarded as a structure built up on particular elementary constituents. As in the statistical thermodynamics, which is constructed on the notion of molecule, our cosmographical models start with galaxy as the elementary unit. It is these agglomerations where the bulk of the ordinary matter is supposed to be concentrated. As we shall see later, there are models/paradigms, which generalize the notion of a cosmic unit, as is the case of hierarchical cosmos, but going below galactic level takes us out of the cosmological sphere proper. These more or less compact units are building blocks of the cosmic large-scale structure. They are well separated, quasi isolated, apart from their weak gravitational interaction and small exchange of the electromagnetic energy. To a large extent, cosmology treats galaxies as physical points, endowed with collective (coherent) and proper (chaotic) motions. This sector of modern cosmology should, however, be treated with caution, in particular when talking about cosmic expansion, space expansion, stability problems etc. Space expansion is surely more fundamental phenomenon (genuine or not), than mere mutual receding of galaxies. Unlike the latter, the former is inevitable entangled with a number of tautological questions, and thus linked with logical conundrums.

### 3.5. Phenomenological, observational and theoretical stabilities

It appears a common experience that small systems change fast, large slowly. This assumption proves correct when dealing with cosmos, but our universe contains actually parts, which are governed by disorder, as the case with Cosmic Background Radiation (CBR) shows. But more crucial situations in that respect are cosmic phases, according to some paradigms, like the inflatory stochastic scenario, when the universe undergoes a phantastically rapid global change. If the effect proves (whatever it means) genuine one,
the difference between small and large might appear irrelevant in this respect.

But more important for our consideration here is new observational evidence, which may change drastically our image of the universe. The case in point is the recent discovery of the cosmic acceleration, which has forced theoreticians to reformulate their models, if not paradigms.[2] From the philosophical point of view, the concept of an ever-expanding universe is sufficiently disturbing for a rational mind, that the additional acceleration makes the case even more astounding.

As for the theoretical, speculative side, the story of resolving Olbers' and Seeliger-Neumann's paradoxes illustrates the best how a small change in the human mind (software) affects drastically our picture of the cosmos. In 1922 Charlier published his famous paper on the hierarchical cosmos, which resolved nicely both paradoxes, putting heavy weight on the fractal model (as is called now) as such. The same year Friedmann published his kinematic solution to the Einstein-Hilbert equation, with expanding universe, which resolved equally well both paradoxes too. Having been based on a powerful physical theory, that of General Relativity, and in particular after Hubble discovered his famous law regarding the galaxy recessions, Friedmann's model pushed Charlier's concept into oblivion (to be recovered half a century later). Charlier's model was essentially static (but see, e.g. Joyce *et al*, 2000; Grujic, 2004) with inhomogeneous, but isotropic galaxy distribution, whereas Friedmann's model assumed both homogeneous and isotropic, but evolving cosmos. One can hardly imagine more drastic change in our vision of the universe we live in, which occurred within 7 years only. It is this possibility to contrive radically different models that we call epistemic cosmic instability.

### 3.6. Saving phenomena

Accepted as a genuine scientific discipline, cosmology is bound to comply with the same rules as the rest of sciences of nature. In particular theoretical predictions, or descriptions, must be in accordance with actual empirical evidence. Since one can hardly expect that a general *ab initio* theoretical scheme have a correct counter part to every new observational data, cosmologists are forced to adjust their models to new empirical situations. These adjustments are generally of various natures. One either makes small amendments to the overall picture of the cosmos, or is forced to intervene at a rather fundamental level of his theoretical construction. Another way out of a mismatch between the paradigm and the empirical reality is to introduce new, extraneous elements into the model, under the proviso they do not contradict the original structure. We shall quote examples of these modifications, aiming at saving phenomena (φαινόμενα σώςειν, as the ancient Greeks would say; see, also, Duhem, 1993).
.

The redshift phenomenon, interpreted as indicating galaxies recession, was a fatal blow to a number of current cosmological models. As mentioned above, it suspended for a long time the hierarchical paradigm, and at least marginalized Steady State paradigm due to Bondi, Gold and Hoyle. The former did not succeed in incorporating the concept of an expanding universe (but see Grujic, 2004), while Hoyle with coworkers did try to accommodate the static paradigm to the newly revealed evolving universe. His ideas about matter creation, via the so-called C-field postulate, necessary to maintain the central point of the static paradigm, removed the main defect of the model, but at the price of loosing the original elegance and simplicity of the static paradigm

Hoyle and Narlikar model sacrificed one of the most sacrosanct laws in physics – the energy conservation, which has never been found violated under ordinary (laboratory or otherwise) circumstances. But that the old dictum *Quod non licet bovi licet Jovi* (*sic*) might be helpful was rediscovered recently, on the occasion of the newly "discovered" dark energy, necessary for justifying the flat universe paradigm (see, e.g. Barrow, 2002).[3]

Another instance of calling θεος απο μεχανες (*Deus ex machina*) was the inflationary phase of the variant of Standard model. Assuming at the same time hypothetical quantum field that pushes the exponential expansion (ontological aspect) and violates relativity constraint on the speed of transmitting signals (c) (epistemic aspect), this mechanism enables one to explain hitherto puzzling empirical evidence of the homogeneous universe (among other things) (see, e. g. Collins *et al*, 1989). But the very fact that introducing into game a relatively simple mechanism, which resolves several questions, does not guarantee any reality of the phenomenon invoked. The physical mechanism, described in broad terms as the present situation is, with a number of hypothetical attributes, remains of a tentative nature, whose heuristic value may be appreciated, but remains of the hypothetical (historical) nature nevertheless. In fact, there have been proposed a number of alternative models, which do not involve inflatory scenario, and which pretend equally convincing resolving the same cosmological puzzles. We shall return to this particular issue later on.

### 4. THEORETICAL ASPECTS

As mentioned before it appears very difficult, if not impossible, to set up a general theoretical framework for modeling cosmological paradigms, independent of the latter. What is tantamount to saying that a deductive approach seems doomed to failure. This leaves little space for contriving complete, self-sufficient theories, which describe the structure and (eventual) evolution of the cosmos. But before entering these tantalizing, logical conundrums, we illustrate the point by discussing briefly two most general physical theories we know at present and their application to the cosmology as such.

General Relativity (GR) was constructed to deal with gravitating matter, without any particular restriction. But after the Quantum theory was formulated physicists became aware of the question of the applicability domain of the new theory. With advent of the quantum field theory it became clear that GR concerns macroscopic and megascopic systems only, or equivalently not too strong gravitational (or any other) field. It remains relevant to the cosmology proper, that is general description of medium and large-scale cosmic structure, but as one follows homothetic transformations of the space involved down to the Planck's scale, GR ceases to describe the physical reality (see 3.3).[4].

The story of Quantum Mechanics (QM) follows along similar lines. It has been construed to solve particular microscopic phenomena, which involve what we now know as elementary particles and fundamental fields (including, presumably, gravitational one), and has been tested within the microscopic domain to the unlimited accuracy. But, as it happens with other mental endeavors, the original theoretical construction was pushed to wider and wider domain of application, far beyond the original one. The rationale for this broadening of the original scope lies in the conviction, albeit tacit, that QM is essentially an epistemic entity, independent of any underlying property of matter (ontological aspect). Since these properties are empirical facts, the claim of an unlimited universality of QM amounts to asserting that the theory is model independent, almost of logical nature. One often meets assertions that QM is applicable to any physical system, in particular macroscopic, even celestial bodies, without any concrete proof of this universality. True, one usually resort to practical difficulties in describing large systems with many "elementary particles", like electrons, protons, neutrinos etc, but our inability to carry out practical calculations in this domain does not entitle us to claim it is possible in principle. Inability does not prove anything beyond inability. This should be kept in mind when attempting to describe Universe by Schrödinger's wave function, as some cosmologists do. QM is a construction of an external observer, endowed with mind (to contrive the theory) and senses (both proper and artificial ones) to test its predictions (or reality, in Einsteinian sense). Attempts to interpret and/or amend QM without observer have not yet provided convincing arguments against Bohr's original interpretation. Thus, treating Universe as an isolated, unique system appears in variance with the very assumptions on which QM was built up. All essential ingredients of the Quantum theory, like the probabilistic accounting from the experimental results (in the broad sense of the concept of experiment), etc become meaningless. Of course, playing with multitude of universes, like playing with multitude of angels on the needle top, is legitimate, from the logical point of view, but little relevant to a theory of serious scientific pretensions. One can not escape comparing some specific radical interpretations/modifications, like Everett's many-world concept, with Einstein's concept of a limited, but borderless Universe.

An essential feature of the Standard (Copenhagen) interpretation of the (phenomenology of) QM is that any experiment consists of a microscopic object and macroscopic (measuring) apparatus, including the very observer. Just how one can imagine the universe as an object of an external observer/experimentator appears a crucial question when manipulating the wave function of the Universe.

Broadening the scope of quantum theory is a legitimate theoretical attempt, but it begs for justification from the existing experience in using QM at the laboratory level. This broadening may take various forms. The most obvious and common is a direct application of the existing formulation, via Schrödinger's equation, for instance. Apart from the very formal structure, QM contains the fundamental constant $h$, which couples it to the physical reality, and as such appears external to the mathematical formalism involved. One is tempted to introduce larger values to this constant as he goes to larger cosmic space, proper to the larger cosmic subsystems, as the case with hierarchical paradigm is, for instance. Thus one may postulate a hierarchy of cosmic Planck's constants $h_s$, depending on the scale parameter $s$ (see, e.g. Grujic 1993 and references therein), with corresponding Schrödinger's equations. This procedure, which amounts to quantizing Planck's quantum (albeit in the opposite direction) might be relevant to treating various levels of cosmic significance, but is hardly a rout to solving the problem of the wave function of the entire Universe.

GR and QM are two most general physical theories we have today. But are they relevant to cosmology in the broadest sense? The answer depends on the particular paradigm we have in mind when posing the question. We know that the very gravitational force, which is the principal ingredient of GR, can destroy the structure of a limited cosmic subsystem, like galaxy(*via* black hole formation, for instance), to such an extent that QM (more precisely, quantum field theory) must be invoked instead to describe the properties and behaviour of the subsystem under most extreme conditions, like an enormous mass density etc. These circumstances remind us that the formal structure of a theoretical construction is not sufficient to consider it a physical theory, without determining its domain of application.

The question of the physical domain of a theory (ontological aspect) appears as a specific instance of a more general issue, that of a completeness of a theoretical constraction (epistemological aspect). The latter has been considered by a number of authors (see, e.g. Hon, 2004). Human eternal strives to find a universal receipt, from stone of wisdom to "theory of everything" (TOE), have always ended in disappointing failure. One of reasons for this lies in limited empirical evidence we can have about the physical reality (ontological aspect), what results in our inability to cover with theoretical constractions a radically new experience. The case in point is cosmology, since we shall never have an opportunity to experience the universe in a direct way. It is this aspect that makes cosmology beyond the deductive approach. But more serious limits are imposed by our mental constractions themselves, as Gödel has so effectively demonstrated in the case of mathematical structure, in particular concerning arithmetic systems (Gödel, 1931). Generally, it turns out that any particular scheme aiming at all-embracing theory is doomed to failure, since it can never be made closed (internal incompleteness). In a less stringent form it turns out that one can never prove that a mathematical system does not contain contradictions. This result is due to mathematical logic and is independent of our empirical experience whatsoever. As Hon argues in his lucid analysis one is forced to choose between an all-embracing system which inevitably fails at some instance and a less ambitious theoretical construction with a limited, well-defined domain. He demonstrates the point referring again to Gödel, but not to his famous proof of the essential incompleteness of mathematical systems, but to his less known, but important contribution to the cosmology itself (Gödel, 1949a,b).

By solving Einstein's GR equation for a class of general rotating cosmos models, Gödel found a solution which deprives the time of a unique direction, thus enabling closed time-like curves to emerge. This turned out to be one of possible arrangements for time-travel, a concept deprived of real (physical) meaning. By doing that Gödel demonstrated that Einstein's GR appears overcomplete, allowing for meaningless effects.

Gödel demonstrated that mathematics, as a formal system, is inexhaustible. Although there is no (yet) an analogous proof that the same applies to physics, in particular to cosmology, it is reasonable to assume that science of Nature is not in a different position. Hence, Platonic world of pure notions need not necessarily be more perfect that the real physical one. What makes our distinction between the universe and Universe merely semantic one.

## 5. MEGACOSM, MACROCOSM AND MICROCOSM

Unlike Plato, whose scientific affinity was shaped mainly by mathematics (in particular geometry), Aristotle ideological background was determined by his biological education, family tradition that could be traced back to Hippocrates. It was this background that explains his remark that the whole is more than a sum of its parts, for this is exactly what distinguishes living organisms from the inert matter. But what about the opposite case, are there instances when the parts are determined, at least in parts, by the whole they belong to? In the case of physical systems cases are known where this sort of holism should be accounted for. Here we consider some instances relevant to cosmology. We shall distinguish, somewhat arbitrarily, three levels of the physical reality: microscopic (essentially quantum mechanical), macroscopic (everyday experience) and megascopic (cosmic) ones. We shall correlate phenomena at a lower level with the properties of the next higher and second higher levels.

The first phenomenon relating micro to macro levels we mention is the so-called Casmir's effect (see, e.g. Lambrecht, 2002), from the quantum theory of radiation. In short, it forbids atom radiating electromagnetically if the macroscopic environment is not adjusted to the emission. In fact, the effect appears an extension of Bohr's postulate of his Old Quantum Theory, which asserts that an excited atom radiates if and only if the frequency of the electromagnetic field satisfies a relation imposed by both initial and (possible) final state of the atom. Casmir's effect adds to these two microscopic boundaries a third, mesoscopic one, demonstrating clearly that quantum transitions are not of a "ballistic nature", but involve both initial and final states. This sort of "teleological aspects" of physical phenomena is thoroughly discussed by Price (1996). The same author analyses in great detail another possible, intriguing coupling between micro- and macro-levels from one side and the universe (Price, 1996). It concerns the Wheeler-Feynman's (classical) Absorber Theory (Wheeler and Feynman, 1945), whose essential part assumes a correlation between emitter's retarded and absorber's advanced potentials. We shall not dwell her on the Price's arguments against the link between (microscopic) emitters and cosmic (megascopic) absorbers, but just quote the original idea of coupling microscopic phenomena with cosmic response. We mention here another two proposed couplings that involve megascopic and lower levels of physical reality.

In his critical reexamination of the Newtonian mechanics Ernst Mach conjectured that the property of inertia of material bodies stems from an interaction with the rest of the universe (see, e.g. Hon, 2004). Since the gravitational force appeared the only universal interaction, with long distance range, it was natural to assume that it is the gravitational interaction, which gives rise to inert mass of a body. At the same time it was a hint, albeit implicit one, at explaining the empirical fact of an equality of the gravitational and inertial mass. The latter would form the content of the Einstein's equivalence principle, the cornerstone of his GR theory. Also, it was Einstein who dubbed Mach's conjecture *Mach principle*. Einstein hoped that this principle would be an essential ingredient of his GR theory, or better that it would be derivable from it. This programme, or hope, never materialized, as we know. What Gödel showed, by deriving a particular structure of the universe, but starting from space-time metric compatible with GR, instead the other way round, was that Mach principle was not an integral part of GR. In his solution time appears without a unique direction, what is tantamount to depriving universe of its cosmic time. Time is still all pervading entity, as with Newton, but is deprived of "rigidity" and interacts with matter. This choice places time on equal footing wit the space, making the whole four-dimensional manifold more "homogeneous" in this respects,

Finally we mention a recent discovery that it is not possible to define the S-matrix within string theory, if the cosmos appears accelerating and has therefore an event horizon. The latter implies a Hilbert space with finite dimensionality, whereas the string theory can not define observable unless dealing with an infinite number of space dimensions. The local observer appears incapable to isolate scattered particle, what makes the scattering matrix indefinable. Since the string theory aims at the ultimate goal of the quantum field theory, i.e. to be "theory of everything", this discrepancy appears a very serious one, in particular in view that it is the theory that tries to treat the universe as a quantum object and pushes the cosmic time beyond the Big Bang (see, e.g. Gasperini and Veneziano, 2002 and references therein)

We have dwelt at some length on the correlations micro- macro- megacosm, for these appear crucial for establishing self-sufficient, all-embracing theoretical schemes of cosmological interest. Interdependence of parts and the whole (ontological aspect) makes an application of deductive method very difficult, if not impossible. On the other hand, *circulus viciosus* which arises inevitably in trying to formulate an overall theoretical scheme (epistemological aspect), need not necessarily be interpreted as a weak point and may be regarded as a sign of self-sufficiency, an ultimate goal of any all-embracing construction.

## 6. THE QUESTION OF UNIQNESS

As stressed before, Universe is not just the largest object we may contemplate about; it is unique, by definition. In order to illustrate the epistemological consequences of this cosmic attribute, let us conceive a quasi-inductive procedure of attaining the Universe as an object of study. Let one start with a finite part of the universe, say containing her immediate cosmic surrounding, like a local galaxy cluster. By broadening the boundary of consecutive ever-larger subsystems one obtains a series of subsystems and may expect to reach the Universe in a suitably defined limit. But the limiting subsystem (or better system) is not a member of the series. It is, by definition, a unique totality of a physical reality. As we mentioned before, this uniqueness makes the universe out of reach of QM, at least within its standard (probabilistic) interpretation. Now the question arises: Is it possible to conceive a totality of physical reality, both in ontological and epistemological sense? Contrary to the assertion by Sherlock Holms that whatever puzzle a human mind may contrive another mind can resolve, in the case of cosmology the following statement may be made: Whichever picture of a unique Universe one constructs, there will be another model which spoils the proposed construction. In the case of cosmology, this destructive attempt appears in various disguises, mainly under the name of *multiverse*, as opposite (or generalization) to the standard notion of Universe. We shall consider cases of proposed multiverses (here plural is used in lexical sense) later on. Here we shall analyze possible meanings that may be attached to this notion.

We start with lexical aspects. For that purpose it is more convenient to use term world instead of cosmos, as will become clear immediately. Hence, our multiverse paradigm transliterates into *many-world* concept, as used in an interpretation of QM, as we shall discuss later on. Term *world* appears homonymous in many languages, in particular Greek, English and French (and Serb, for that matter). In French, it refers first to a particular small community, like family, class etc, (*tout le monde va bien* – all people in the family are well). Then, it may concern a wider social environment, meaning public opinion (*devant le monde* – in public). Humanity is *le monde*, too, as the name of daily *Le Monde* implies. It is easily extended to other, extraterrestrial civilizations. Finally, the term may refer to the entire conceivable universe. Similarly, in Greek, cosmos means first human community (*elefteros cosmos* – free world), planetary system (originally, cosmogony referred to the formation of our Solar system only), then extraterrestrial (and extra-Solar) civilizations, to comprise eventually the entire possible universe. Homonyms are common in all languages, of course, but we stress here the hierarchical ordering of (usage of) "world", referring to ever-broader systems, as just illustrated. It illustrates, at the same time, the order in which human developed their worldview, or how the concept of the universe (epistemic aspect) followed the empirical evolution (ontological aspect).

Distinguishing physical systems involves two notions: distinct and isolated. Two identical, ideally isolated systems are of little interest to us, for it reduces to a mere numerical plurality. Somewhat less trivial case is that of Anaxagoras (Grujic, 2001 and references therein), where various hierarchical levels are linked by the homothety transformation. Generally, one can obtain a multiverse out of Universe in various ways. We shall discussed particular realizations when considering various cosmological paradigms, and here we confine ourselves to a number of conceptual choices.

The simplest way to get many worlds is to partition the existing universe into weakly interacting, quasi-isolated subsystems. The only exchange between such subsystems would be at the informational level, like via electromagnetic radiation (visual evidence). We observe distant galaxies, cluster etc, but even an information exchange (and any other possible contact) is out of question. Further, we may envisage distant subsystems, which are in no interaction with our and other parts of the universe, but are nevertheless correlated with them. This case is exemplified by quantum distant correlations, as illustrated by wave function entanglement of EPR type (see, e.g. Laloe, 2001, and references therein). This case is present in the inflatory scenario, which presupposes an initial strong mutual interaction, with issuing correlations preserved after inflatory loss of the causal relations. One might even assume the possibility of existence of a set of noninteracting subsystems, totally mutually isolated, but with a common "wave function" (or any other function of state), which account for their existence and reveals that fact by statistical properties and their consequences on any of subsystems. Obviously one may speak of a "system" in such a case in statistical sense only (even term "cosmos' would be inappropriate).

One of bizarre concepts of multiverse is that due to Everett (see, e.g. Barrow and Tipler, 1986), with his many-world "interpretation" of QM, better to say manipulation of Schrödinger equation. The idea is that after any quantum measurement (event) the world multiplies into many branches and each of newly formed universes loses (causal) contact with the rest of the physical reality. The idea rests on the concept of the wave function of the Universe as a unique, isolated entity, the concept that has no support in the original derivation of the Schrödinger equation.

Generally, there is no way to prevent somebody to multiply the universe one conceives, either in spatial, or temporary sense, or both. The last case in point is the so-called stochastic inflatory scenario (paradigm), as envisaged by Linde (see, e.g. Linde, 1987). We shall consider these models later on. Here we just note how evasive the concept of absolute totality of physical reality is, like other notions not subject to the empirical evidence, like that of infinity, eternity etc.

## 7. COSMOLOGICAL MODELS – A QUESTION OF INSPIRATION

Verification of a theoretical construction may take various forms, first of all by empirical (possibly experimental) evidence. In the case of far stretching theoretical models, like many of proposed cosmological ones, when the empirical evidence is lacking, other means of estimating credibility to be assigned to a particular construction must be found. It is here that epistemological considerations appear invaluable tools for valuing merits of ideas, assumptions etc.

### 7.1. Anthropocentric inspirations

One of the most reliable criteria for assigning credit to a cosmological model is the rationale on which the idea rests. Cosmological models, even paradigms, stem from various motivations that have inspired their authors. Often, it is by looking at these inspirations that one may discard from the start some (i)rational construction. One need not bother about finding physical (causal) links between a celestial constellations and human destiny, for instance - it would suffice to look at the way astrologers contrive their recipes. Similarly, if we find a mythological inspiration, or a religious one, we know we are far from a hard science and are entitled to be suspicious if the model appears too speculative. Some models, however, appear of a mixed type, including both rational and nonscientific elements. The case in point is Anaximander's cosmology, the first rational one, making the bridge between a mythopoetic and scientific picture of the cosmos. Anaximander conceived the cosmos of his time, that is as implied by the observational inference, as Earth shaped as a drum and three concentric spheres around it: astral, with Moon and Sun, respectively. The ratio of the corresponding diameters was 1:2:3. The revolutionary step was the assumption of an autonomous celestial system, with no external support, imbedded into the cosmic space. The puzzle remained, however, what was the meaning (or rationale) for the particular choice of this series, both in number of terms and in figures adopted. Four possible inspirations have been invoked by various authors: (i) mythological, (ii) astronomical, (iii) architectural and (iv) sociological. In his recent article Naddaf (1998) argues for the last one, attributing the numbers to three social classes of the Greek polis: aristocratic, middle (urban) and peasant (poor) class respectively. His concern was, apparently, the social harmony (read stability) of the polis of his time, and he might have argued for it *via* cosmological model which stands at a lofty position with respect to "Earthly" things and thus can serve as a guiding principle, at least. Naddaf argues that these celestial spheres stand in the same proportions as the distances of those social classes stand relative to the *agora,* the focus of the polis.[5]

Generally, we remark that the scientific methodology may be eventually reduced either to tautology or to analogy. Cosmological thinking often resorts to the latter, following the ancient formula: microcosm ~ macrocosm.[6] (Biblical Leviathan myth, as used by Hobbs, belongs to the same category of thinking). Microcosm as an *ansatz* may be regarded as a specific language, in Barthes' terms mentioned above, in cosmological matters, just as macrocosm is treated as the (classical) language in Bohr's (Copenhagen) interpretation of QM (Laloe, 2001).

It is interesting to notice how much some modern cosmological models resemble ancient mythological solutions, and even more some esoteric teachings, like that of Kabbala and its concepts of *Ain Soph* and *Zimzum* (see, e.g.. Masson, 1970). Since many of European thinkers and scientists (*y compris* Newton and Leibniz) were members of various esoteric societies it would be of great interest to examine influence of these esoteric teachings on the rational systems we owe to them, especially cosmology. But it would lead us beyond the scope of the present work.

### 7.2. Anthropic principles

These principles are intermingled with so-called selection bias, teleological bias, but appear not identical with these. They are hardly principles and anthropic bias (see, e.g. Bostrom, 2002; see, also, Cirkovic, 2002) would be more appropriate to use, but since this terminology is in common use now, we shall retain it here (see the classic work on the subject by Barrow and Tipler, 1986). There have been formulated several of them, with different both content and nature. The so-called *Weak anthropic principle* is actually an observation that whatever we think about the evolution of the universe, the fact that we exist (as observers) must be accounted for. This sounds as tautology,[7] but in fact the statement is not deprived of power of retrospective prediction, as the famous case of Hoyle's inference of a nuclear reaction demonstrates (see e.g. Barrow and Tipler, 1986). The logical content of the assertion is that the past has been determined by the present, contrary to the usual formulation of the causal relations (and intuitive expectation).

The *Strong anthropic principle* goes one step further asserting that Nature is endowed with purpose, in view of the existence of intelligent beings like ours. This evidently teleological thesis, with a strong "spiritual" (not to say religious) flavour might be taken as another aspect (or consequence) of the pantheistic worldview, but we shall not dwell on it here and direct interesting readers to the classic work by Barrow and Tipler (1986).

### 8. COSMOLOGICAL PARADIGMS AND MODELS

*Never run after a bus or cosmological model,*
*for there will be next one in a few minutes.*

Professor of French history, Yale

As mentioned before, the number of cosmological models, even paradigms, is so large, that few common features are found among them. This makes a general discussion of the epistemic questions in cosmology difficult to exercise in generic terms. We shall proceed, therefore, considering a number principal cosmological paradigm that one finds on the market, both historical and actual.

## 8.1. Pre-Socratic cosmology

### 8.1.1. Abderian universe

This cosmological paradigm, conceived by Leucippus and Democritus (e.g. Kirk *et al*, 1983) stands at the beginning of a series of models, which arrives to our time under various names (see Grujic, 2001 for discussion on this cosmological line of thought). It is the first cosmology in proper sense, as we conceive the subject today. Abderian cosmology is based on their atomistic hypothesis and thus represent the first theoretical constraction that links microcosms and macrocosms (more precisely, mega cosmos), albeit in qualitative terms, the more rigorous scientific level achieved only recently in modern cosmology (see, e.g. Collins *et al*, 1989). The concept involves infinities, both at microcosmic (atomic) and mega cosmic (cosmological) levels. As Lucretius informs us, an argument against a finite universe was that in the latter case all masses would concentrate at its centre (Lucretius, 1984-1997), implying that gravity affects all bodies, not just heavy ones (Russo, 2004). The first refers to an infinite, countable number of the primary cosmic constituents (atoms), while the spatio-temporal attributes of the universe on large scale are supposed unlimited. Within this unlimited, eternal universe an unlimited number of cosmoses are present, where cosmos as a constraction is applied to systems (hence ordered parts of the universe) like our planetary cosmos (we could designate this cosmological level as macrocosmic, situated between atomic and proper cosmological ones). These cosmoses may be inhabited, possibly by intelligent beings like ours, and are subject to changes, with ones perishing and others coming into being. Though atomists never state it explicitly, atoms are eternal and hence the universe they are basic constituents of, appears consequently eternal too. It appears purely materialistic concept of the world, explicitly atheistic, what proved almost fatal for its fate, being ignored by the subsequent philosophers and damned (and ignored) by the Christian church. It possessed two appealing features: the concept of an eternal overall existence (ontological aspect) and evolving partial structure (dynamic aspect). The question of "initial conditions" (initiation problem) and boundary limits are solved automatically, being superfluous within this paradigm. Hence, it appears that the first scientifically conceived universe was at the same time the most general one, adopting, albeit implicitly, the perfect cosmological principle (better to say the principle of *isonomy*, as a more general concept used by Greeks at the time (e.g. Vizghin, 1989).

Abderian universe reappears during the European history many times disguised under various names and in a number of variations on the theme, as we shall briefly mention. Besides Epicures and Lucretius, mentioned should be made of Kant's cosmology, as exposed in his famous treatise (Kant, 1968). Eternity, an infinite extension, evolutionary (more precisely, cyclic) changes of ordering etc are present in Kant's cosmology too. In addition, Kant incorporates into his scheme the concept of hierarchy, too (see, e.g. Mandelbrot, 1982). Marxists adopted Abderian cosmological paradigm, too, as exposed in Engels' *Anti-Duhring* (1878). The more recent concept of the so-called Steady-state universe (Gold, Bondi, 1948; Hoyle, 1948) (see, e.g. Narlikar, 1977) belongs to this Abderian paradigm, too, starting explicitly from the perfect cosmological principle.

Abderian paradigm solves, albeit implicitly, major cosmological puzzles that have been occupying human minds from the beginning of prehistoric times, that may be cast into questions: Who are we, whence we come from and where are we going? As the *homo sapiens* reached the historic times, the questions took the form of a less anthropic disguise: What is it all about? To Democritus the question of the Beginning and (eventually implied) End was apparently devoid of meaning. This solves, in passing, the question of the First mover, Demiurg etc, eschatological dilemmas, and the host of other puzzles one might associate with the fundamental cosmological issue. If Hesiod writes about gods in his allegorical poem *Theogony,* some cosmologists (in particular those within a religious approach) call for the external help in the form of god(s), Abderians (and their followers) do not feel obliged to resort to any entity external to their rational concepts. Their cosmology appears complete, albeit in qualitative terms only. It is also "structurally stable", that is minor amendments and modifications do not destroy the overall picture of our ultimate physical reality.

If (macro)cosmoses come into being or vanish, it is only in the structural sense, as the very meaning of "cosmos" implies. Absolute becoming and perishing appears impossible, but not because of a physical principle or the like, but since it is devoid of a logical content. Hence, the "problem" is not of scientific nature, but belongs to our way of thinking, to the very logic our mind conceives and makes use of. Nothingness can be neither proved nor disproved and thus does not constitute the subject of a scientific theory, in Popperian sense.[8] Atoms are changeless, they are eternal and such is the universe. These features of the Abderian cosmology have remained the constant ingredients of European scientific culture until the present day.[9]

### 8.1.2. Anaxagorian cosmology

Abderian atomic hypothesis and the resulting cosmology was a response to the Eleatic challenge to plurality (see, e.g. Kirk *et al*, 1983). Another, alternative solution to the problem how to conceive a discrete, infinite universe, was offered by Anaxagoras from Clazomenae (Diels, 1974, Kirk *et al*, 1983). By designing his hierarchical cosmos, Anaxagoras was able to construct an infinite universe out of a finite quantity of matter (see, e.g. Grujic, 2001).[10] His paradigm has reached us under the modern term of a fractal cosmos (see, e.g. Baryshev and Teerikorpi, 2002, and references therein). The central constraction of his cosmology is the thesis that in the physical world everything is contained in everything, or *vice versa*, everything is a part of everything. The constraction appears an inevitable result of an assumption microcosm ~ macrocosm and the requirement that the latter be infinite. It is, essentially, an atomistic concept, too, but not at material level – what is indestructible is the form, in Aristotelian sense. The same form repeats at each subsequent level of reality, toward smaller as well as toward larger (sub)systems. Notions of system and subsystem appear relative, they are interchangeable. There is no reference level, no "natural length". Performing a *homothetic* (discrete) transformation, which just shrinks (expands) dimension of an object (*scaling* operation in modern terms), one passes from one level to another, *ad infinitum*. Formally, this is accomplished by the so-called renormalization operator **R**

$$H_{n+1} = \mathbf{R} H_n \qquad (1)$$

which transforms Hamiltonian *H* at *n*-th level to another at the next one at *(n+1)*-th level. These Hamiltonians must be of the same structure, though in the cosmic reality not necessarily with the same parameters (Combes, 1999).

At an undefined remote instance an agency, whom Anaxagoras named Nous (Νουζ - Mind) started differentiation of the primordial mixture of seeds (σπερματα). The Nous apparently played somewhat ambiguous (undifferentiated) role of (physical) principle and supernatural agency at the same time. Aristotle later dubbed Anaxagoras' seed *homoeomere (ομοιομερη)*, somewhat ironically (entities that are parts of themselves, plainly "that have same parts").

What is the status of the hierarchical paradigm in practical terms now? The model was revived by D'Albe (1907) and Charlier (1908, 1922), to acquire a permanent (though not much prominent) place on the cosmological arena in the second half of 20th century. It is currently known under the name fractal cosmology and the empirical astronomical evidence confirms that the visible universe appears fractally structured, up to some large cosmic distance (see, e.g. Sylos-Labini *et al*, 1998), with the fractal dimension close to *D = 2*. From an epistemic point of view fractal paradigm, as distinct from all other currently in use, should not be discarded, as argued by a number of researchers in the field (see, e.g. Ribeiro and Videira, 1998), referring to the Boltzmannian epistemological argument.

While Anaxagoras' many-world structure is clearly discernable when going towards arbitrary small, his cosmology towards megacosmic dimensions (cosmology proper in modern sense of the term) appears much less obvious (Grujic, 2001). In a sense his ideological base bears an implicit contradiction. Anaxagoras is known to have first introduced the concept of homogeneity (now called Copernican principle), by asserting that the heavenly bodies are of the same nature as earthly ones (for which opinion he was banished from Athens), yet he argued for an infinity of worlds. The latter are of the same structure, however (though of possibly different dimensions) and thus this infinity is of a numeric nature. As mentioned before, Anaxagoras' concept of homogeneity (ontological aspect) is based on the principle of *isonomy* (epistemological aspect).

What D'Albe wrote was a science-fiction-like book in which he applied Anaxagorian concept of self-similarity toward both smaller and larger dimensions. Charlier put the idea into mathematical clothes and was able to eliminate both puzzles that bothered astronomers since Newton's time, so-called Olbers' (blazing-sky) and Seeliger-Neumann's (gravitational) paradoxes. These paradoxes illustrate in their part how local properties might be determined by the overall structure of the universe. Since they are observable effects, they may serve to select among the existing models those that "save the phenomena" (we shall return to this point later on).

Modern cosmology has abandoned fractal concept below galactic level, but the astronomical evidence confirms hierarchical structuring up to supercluster level. The concept appears inapplicable on the universal scale, first of all due to the quantum nature of the microcosm, though in principle it might hold for large-scale structuring up to yet undetermined scale (see, e.g. Sylos-Labini *et al*, 1998)

Before we leave hierarchical paradigm a few words on the methodology are in order. A number of authors arguing against the model used a convincing interpretation of the empirical astrophysical data available to show that the actual dimensionality of the visible cosmos is *D = 3*. However, as demonstrated by Pietronero (1987), the statistical procedure employed was erroneous. Since it was based on the assumption that there exists a nonzero average cosmic matter density ρ, the conclusion arrived at appears a direct consequence of the initial premise. The hierarchical model implies, on the other hand, a zero cosmic density and if it is accounted for, the data observed turn out compatible with the (multi)fractal structure.

 This illustrates clearly the impossibility of making use of a deductive method in dealing with cosmological issues, the partial assumptions depending on the overall conceptual structure. The same applies to the relativistic cosmologies, where various observables, like distances, appear dependent on the model adopted (see, e.g., Ribeiro, 2005)

### 8.2. The Standard model

As the name implies this paradigm is widely (though not absolutely) accepted by the modern cosmologists (see, e.g. Barrow, 2002, for a recent, popular account). It is usually designated as Friedmann-Robertson-Walker model (FRW), as a dynamic solution of Einstein's equation in GR, endowed with Robertson-Walker metric. Dynamic appears determined by the parameter *k = -1, 0, 1,* which determines the curvature of the universe in the four-dimensional space-time manifold. The first value defines an open universe, which expands infinitely in space and time, whereas the last corresponds to the so-called closed universe, which expands up to some maximum value for the scaling parameter *S(t),* which determines intergalactic distance. The border case (*k=0*) implies an ever-expanding universe, which possesses zero total energy (flat, Euclidian cosmos), with the expansion rate slowing down to zero in the limit t$\to \infty$. The latter model is a favourite one amongst the present-day cosmologists, in particular those sticking to the so-called inflatory scenario (see, e.g. Collins *et al*, 1989, Ryden, 2003). The common feature of these models is the start of expansion from a point-like singularity, or at least from extremely small, dense and hot state, the so-called Hot Big Bang (HBB) model. We shall first consider some general features of these models, which are more of a philosophical nature, before discussing specific epistemological questions regarding the Big Bang paradigm.

### 8.2.1. Closed universe, oscillatory model.

It is assumed that at the start of the expansion the overall entropy was low, increasing as the expansion proceeds. The expansion defines a global, cosmic time (arrow), whereas the change of the state of the content determines the thermodynamic time (arrow). As the universe reaches its maximum spatial extension, it starts shrinking, recollapsing eventually into state corresponding to the initial one. One thus speaks about a cosmic cycle, or more formally, about the cyclic time. Since the initial/final state is assumed chaotic, no information is passed from the previous to the subsequent cycle and one is entitled to talk about cosmic beginning and end, respectively. This picture thus defines a many-worlds scenario, not in spatial but in temporal sense. A sequence of universes thus appears, with a specific universe coming into being and perishing into the "final" Big Crunch (BC) singularity. This is essentially an Empedoclian scenario, which ensures eternity via an infinite series of cycles. But we note that though one finds a forerunner for this model in Pre-Socratic Greece (as one finds them for almost any other cosmological model), the concept of "eternal return",[11] is more proper to Indian cosmic ideology (see, e.g. Masson-Oursel, P. and L. Morin, 1978),[12] than European one. True, Democritus, Anaxagoras and some other Pre-Socratic do speak about perishing and coming into being, but these phenomena refer to macrocosm, not to the entire universe.

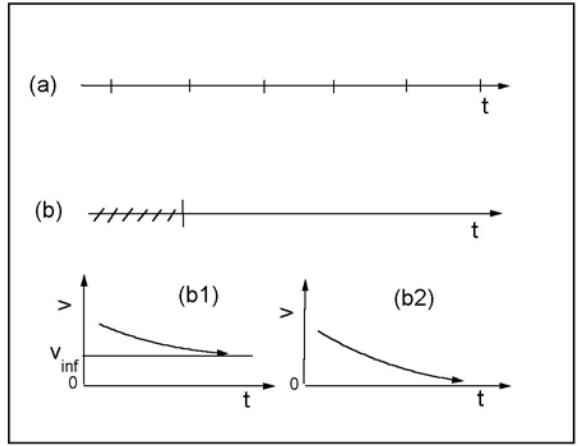

**Figure 1.** *(a) Closed (cyclic) Universe; (b) Open Universe. (b1) Flat (Euclidean) space; (b2) Hyperbolic space. (schematic, see text).*

Formally, one can represent the time evolution of cyclic universe by a series of open intervals on an infinite line, as shown in Figure 1a, with cosmic time stretching from $(-\infty)$ to $(+\infty)$. Points indicated by vertical dashes correspond to initial/final singularities, which are beyond our evidence (and even apprehension) and are to be excluded. Whether the intervals are to be disconnected or not, depends on the physics of the singularities, better to say on the information content and fate within these extreme regions. What appears a matter of controversy at present (we shall come back to this point later on).

### 8.2.2. Open universe model.

Temporal evolution of this model is presented schematically in Figure 1b. The initial singularity is indicated, again, by the vertical dash. The segment is open to the right (infinite future), whereas situation before the Big Bang appears undefined. If the total energy is positive, the expansion velocity remains nonzero in the limit $t \to +\infty$, as indicated in Figure 1(b1). In the case of $E = 0$ the limiting velocity of the receding galaxies is zero too (Figure 1(b2)). This border case appears the most probable according the present day astronomical evidence and we shall discuss it in more detail from the epistemological point of view. Here we restrict ourselves to ideological aspects of this choice.

The model implies, albeit implicitly, the concept of an absolute beginning, and a simple eternal further evolution (*sic*). It has essentially historical character and resembles (if not reflects) in this respect Judeo-Christian world-view. The question of the past, before Big Bang, which is the matter of intensive (theoretical) research now (e.g. Gasperini and Veneziano, 2002), might provide a more acceptable picture of the possible physical mechanism involved, but could not change the basic feature of the model – the uniqueness of the present universe, in the strongest possible sense. Even if the at present mysterious "prehistory" is elaborated (it might turn out that the model is symmetric with respect to the initial Big Bang instance and thus two time instances are present instead of one), the initial question remains as to how could one comprehend a single "historical" event - coming into being and (practically) vanishing of a unique physical system, as the universe appears defined. (We shall address the question of "vanishing of the universe" later on, when considering the concept of physical eschatology).

## 9. EPISTEMIC QUESTIONS

In the previous chapter we considered "ideological" aspects of the current cosmological paradigms, related more to our intuitive feelings than to proper methodological and epistemic aspects of the subject. We shall now devote more space to the latter.

We start with the Abderian paradigm. Its principal feature is the static global state (megacosmos level) and dynamic partial state (macrocosmic level). Since the Steady State model retains, albeit implicitly, the same properties, we address the question of the evolution and the anthropic principles, as pertinent to both ancient and modern models. If the universe is eternal and allows for (evolutionary) changes within its parts, what could one expect from the "final results" of these macrocosmic developments? Could one envisage an infinite duration of such partial evolutions, leading to advanced civilizations like ours and beyond (see, e.g. Ćirković and Bostrom, 2000; Ćirković, 2002). In particular, what's about the so-called Great Silence (as an anticipatory response to famous Fermi's cry, "Where are they?"), and the ETI hypothesis  (see, e.g. Ćirković, 2004 and references therein)?  The most reasonable question one might pose here is a requirement of elaborating in more quantitative terms the problem of macrocosmic eschatology, before going to resolve the problem at the level of the entire universe. It seems desirable to determine the practical meaning of the vastness of the universe, in terms of measurable quantities relevant to the issue, like speed of light (or any other communication means). In particular, what would be the size of internally connected cosmic region, where the information exchange period is commensurable with the span of a reasonable number of successive generations of intelligent beings? Or equivalently, what would be the anthropic horizon determined in that way? But these questions would lead us beyond the scope of the present article.

These and similar questions, still awaiting for answers, despite obvious difficulties  (presumably not insurmountable), do not discard the paradigm as such, as the ideological background of the eternal universe make an pleasing appeal to our intuitive feeling that there must be something eternal as a background to our mesoscopic/macroscopic mortality.

In dealing with Anaxagoras' paradigm, we must distinguish two aspects of this hierarchical concept. First, original one, which is more of a heuristic value and the present-day situation, as evidenced by modern observations. Anaxagoras original idea was a dynamic Universe, which, under the action of Mind (Nous) passes gradually into Cosmos. What is the rate of this process, at least concerning its inward direction, towards the ever smaller parts (Mugler, 1956)? Is thus conceived Cosmos of finite age? What difference does it make when considering the process towards larger dimensions, outwards? What would be the cardinal number of "worlds" in this Universe? Coming back to the universe, a legitimate question is posed as to the consequence of the observed expansion, in particular accelerated one. An attempt has been made to incorporate this sort of dynamics into the hierarchical Charlier's model, albeit in a qualitative manner (Grujic, 2004), but realistic calculations are still to be made in this direction.  On the other hand, fractal model appears well suited for the general case of expanding universe, as shown within Friedman-Robertson-Walker model by Joyce *et al* (2000).

The most interesting and important case is that of the Standard model, which is the subject of the overwhelming number of investigations today. We first consider topologically closed universe, as intuitively most appealing model.

Hot Big Bang oscillatory model is a mixed cosmological constraction, with (most of time) deterministically evolving universe, shuttling between the initial (HBB) and final (BC) states. The latter may be thought of as black boxes, with chaotic content, determining initial/final conditions in an unpredictable way. The intermediate deterministic evolution may be interrupted by one or more stochastic phases, as the case with the inflatory scenario is, but this brings in nothing qualitatively new to the overall picture. One general question is posed when considering this class of solutions to the Einstein-field equation: in which sense the repetitive processes may be assigned to the global (megascopic) and local (microscopic/macroscopic) levels? In particular, what is the connection between the cosmic and thermodynamic arrows o time? Many disputes have been reported in the current literature on the subject (see, e.g. Price, 1996), in particular in answering the central question: when the universe starts contracting (i.e. the cosmic time arrow reverses its sense), what happens to the thermodynamic arrow, does it reverses too? The question bears as much ontological as epistemological significance. Within the classical (dynamics) picture, the answer seems to be positive, if we accept that the underlying dynamics is time-symmetric. This conclusion appears wrong even within the classical picture, since it ignores the essential difference between global changes, governed by the gravitational interaction and thermodynamic, microscopic processes. The first refer to the galaxies as primary cosmic units, the latter to the atomic systems, where nongravitational forces, like the electrodynamic ones, are dominant. Galaxies may recede from each other, but the celestial bodies, like stars, planets, cosmic dust grains etc do not expand themselves, and retain their properties after the general shrinking on the cosmic scales, obeying most general thermodynamic rules. Some authors, however, maintain that the entropy increases during each cycle, with singularities preserving this tendency. Since the cycle periods are proportional to the entropy, they form a series of ever increasing lengths. What implies that going backwards in time one approach a zero-period cycle, what appears tantamount to the very singularity, and the simple repeatitory picture disappears. Thus this scenario appears indistinguishable, for all practical purposes, from the open universe model.

Recently, Steinhardt (2002) proposed a model based on the so-called brane-world, which has a cyclic character. Brane hypothesis maintains that the actual universe is embodied into many-dimensional hyperspace (usually of 11 dimensions), four spatial, one for time and the rest compactified to the extent that this portion is unobservable. All massive and nonmassive particles move in the three-dimensional brane, whereas the fourth dimension is accessible to the gravitational field only. According to this picture, the universe consists of two branes, connected by the gravitating fourth dimension. The brane sectors experience cosmic expansion, as in the ordinary HBB scenario and contraction along the fourth dimension. The mutual approach of the two brane branches results in "universal collision", so-called ekpyrotic phase (Khoury *et al*, 2001). After the bounce the branes separate again and recede from each other, giving rise to a gigantic cosmic oscillator. The model gets rid of the inflation assumption and the dark energy constraction, providing another scenario for the cyclic cosmic dynamics (in fact quasicyclic, see later). Since two branes are causally and otherwise disconnected, apart from the gravitational interaction, this model may be thought of as two-world universe.

The open universe model includes two cases, the flat and the hyperbolic ones. The latter appears less interesting for us, though in the light of the recently observed cosmic acceleration it might again attract attention of the cosmologists.

The flat, Euclidean cosmos is required by the inflatory scenario (Guth, 2002, Steinhardt, preprint). Cosmic, more precisely universe evolution is proposed in many phases. First, the HBB, then the inflatory phase, followed by the normal expansion of the ordinary matter, as we witness today, and then another acceleration, as recently discovered. As we mentioned before, the inflation has been called in to save the phenomena, first of all to explain the flatness, the homogeneity of the universe and finally the absence of the cosmic monsters - gigantic magnetic monopoles. It does do the job, but at the price of introducing a number of entities, which are equally strange and vague. It starts with a small, highly concentrated, homogeneous universe, then stretches it practically momentarily, and then leaves the cosmos to continue with a slow expansion, retaining its initial feature of homogeneity. The scenario overrides, better to say circumvents, the basic STR request – maximum speed, $c$. It does not violate, however, a more fundamental requirement, that of the transmission of the information limit. The universal expansion implies a coherent motion of the receding entities and the fundamental causality requirements are not violated for any subsystem of possibly interacting bodies. In fact, it is the physical (massless) fields that undergo inflatory expansion. And it is these fields that are the most mysterious entities within the entire concept, for their mathematical form is very arbitrary and many particular potentials as functions of the hypothetical field are on the market last decades. The origin of these fields, not to mention their mathematical structure, remains obscure to these days and no advent has been made in elucidating their nature up to now. Hence, it seems that a big cosmological puzzle is to be solved by relying on mysterious entities, which should have acted within an extremely short period compared with the previous cosmic age, let alone the entire age of the universe.

Inflatory scenario was proposed to explain the homogeneity of the universe, thus fulfilling Copernican principle. On the other hand, since we have discovered the cosmic acceleration, it turns out that we live in a singular cosmic time between deceleration and acceleration phases. That it happens to us has perhaps something to do with the anthropic principle, or better the sampling bias, remains to be seen.

### 10. UNIVERSE *VS* MULTIVERSE

Crucial points of every cosmological model is how it treats the problem of Beginning (and eventually End), with accompanying questions of the initial conditions and relating eschatological issue. Abderian paradigm, fulfilling the perfect cosmological principle, is immune to these problems. It is not the case with other solutions, particularly with the Standard model. Generally, an effective theory includes a formal mathematical model and the procedure how this couples to the physical reality. The latter amounts to prescribing initial or boundary conditions. They are put by hand, expressing our choice, which is governed usually by the circumstances or other preferences. The universe, however, is not a toy at our disposition and the choice acquires quite a different meaning compared with laboratory experiments. Religions circumvent this problem by giving mandate to the choice to gods, with provision that their existence (and will) is nonquestionable. This choice is not available to science, and certainly not to cosmology.

There are, generally, two ways out of this problem: either eliminating the creation issue or multiplying the act of creation to an infinite number. The first solution, proposed by Hawking (1982) and Hartle and Hawking (1983) aims at formulating a model which is self-sufficient, in the sense that the very formal structure implies boundary condition, or what is tantamount to this, model in which there is no place for them (cosmos without initial conditions). This solution seemed very acceptable at the time, at least promising, but we have witnessed appearance of a number of other models since then, that the issue took over quite different nature. We briefly discuss one of the most popular, though not universally accepted model, that of the stochastic, inflatory, eternally self-reproducing cosmos scenario due to Linde (e.g. Linde, 1985, Linde, 1998).

The scenario is self-sufficient (ontological aspect), but this self-sufficiency brings in methodological obstacles (epistemological aspect). We start with the latter by saying "we start …". One borrows from the Standard model with inflation the initial quantum fields that trigger and carry the inflation. These fields are subjects to inevitable quantum fluctuations, which have two essential features, required for our purpose. They are random in every respect and are seeds for forming new cosmoses after the inflation is over. Randomness concerns not only their nature, but also the instance of appearance. Hence, the initial conditions are unpredictable and thus are out of our (theoretical) control. If Hawking afforded to defies gods (or God?) (see, e.g. Hawking, 1988), Linde defies, albeit implicitly, Einstein, who at the time forbade gods (or God?) to play dice with Nature. Quantum fields do acquire gods' nature, at least as far as the unpredictability is concerned (the latter being the essence of a free will). Linde's cosmoses thus are born out of these primordial fluctuations which grow in time and turn into worlds like ours, but possibly quite different from ours. They might, in principle, possess entirely different structure, obeying their own physical laws, have their proper dimensionalities etc. In other words, they fill up the entire "epistemological space", fulfilling tacitly the principle that anything that is possible (not forbidden) does exists.

The destiny of these newly born cosmoses is described by the standard scenario from the Standard model. They recollapse into (black-hole-like) singularities, and reappear as new cosmoses and the game repeats *ad infinitum*. There are infinitely many cosmoses, at infinitely different "universal" times, coming into being and perishing, in an essentially aleatoire manner. There in no beginning, no end, only eternal changes and transformations, like a living organism. Whether this being should be called Universe or multiverse, or multicosmos, etc, is a matter of semantics.

Before we leave the subject, two remarks are in order here, First, the usual notion of causality should be reassessed when considering disappearance of a cosmos into the black hole (Big Crunch) and reappearing new cosmos out of this. The first crunch is a cause of the ensuing bang, that for sure. But since the new cosmos has nothing to do with the first one, in any respect, it is hardly possible to talk about causal connections. Second, it is not possible to describe a phenomenon which is, by author's intention (and thus by definition) without beginning (and end). Hence the phrase "We start ..." looses meaning, since it defies the very concept of infinity, implicit in the very paradigm of this scenario. Ontological completeness rules out epistemological self-sufficiently, our reasoning being essentially along a linear path. To put it differently, we have no Archimedean point to start from.

## 11. PHYSICS, ASTROPHYSICS AND METAPHYSICS

*Cosmology is a continuation of mythology*
*by mathematical means.*
P. G.

Every scientific discipline has many levels of research, ranging from fundamental stratum down to applicative domain. The latter is usually of a technical nature, whereas the front of the science contains much of hypothetic, speculative, tentative etc material. This holds particularly for cosmology, as well for the field-theoretical research in the sector of high-energy physics. Since the latter appears an essential tool for investigating the most important segments of cosmological studies, the fact that both disciplines acquire hypothetical nature at the forefront makes their link very complex and even somewhat dubious indeed. It is at the fundamental level that models and theories abound in both fields and one is aware that the overwhelming majority of the models proposed must end in historical archive. Not that any of them appears unconvincing, but looking at the plethora of proposals, many of them being radically different, if not contradictory, one is aware of the arbitrary nature of the theoretical constractions invented. It is this arbitrariness, together with authors' overselfconfidence, that prompted some researchers, *y compris* those engaged in cosmological studies themselves, to make a parallel between the mythology and modern cosmology (see. e.g. Whitrow and Bondi, 1954, Alfven, 1976, Lurcat, 1978, Luminet and Magnan, 1993; see, also, a recent critics by Disney, 2000).[13]

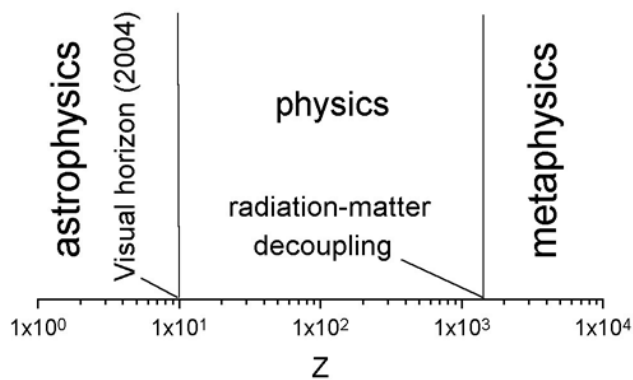

**Figure 2.** *Epistemological zones for Hot Big Bang scenario (Standard model)*

We shall illustrate this aspect of the cosmological studies by considering the Standard model. One can separate spatial cosmic regions by introducing (visual) horizons. Due to the finite speed of light, this division can be conveniently transformed into cosmic time instances, or into the corresponding red shifts, $z$, as we adopt it for the graphical convenience. In Figure 2 we show schematically the cosmic space division into three characteristic regions. The actual visual horizon corresponds to our everyday usage of the term, the region we can observe at this earthly instance. This horizon is pushed away as we construct more power devices for astronomical observations, principally the telescopes. It is within this visually accessible part of the universe that we are able to test our theoretical models. We shall call this part of the surrounding space astrophysical region, which one may treat as a vast cosmic laboratory, not for experimentation, but for observational empirical evidence. We "control" this region by improving the observational tools and the resulting visual horizon changes noticeably within the span of a single generation on Earth. The next characteristic distance is the surface of the last scattering (LSS) of the cosmic electromagnetic radiation, when the latter decoupled from the matter and cosmos became transparent and thus subject to observation by the transmitted electromagnetic radiation, either stemming from luminous sources, or from dark illuminated celestial objects. Beyond LSS a complete chaos of electromagnetic and massive matter is envisaged. Because of LSS this region appears (*sic*) invisible, though we do receive radiation (and possibly even particles) from that "dark vilayet". Chaotic motion prevents any structure to be discerned, but fortunately (by definition) there is no structure to be seen. In principle, one might hope to gain some information from this region by some other physical carrier, like the gravitational field (see, e.g. Steinhardt, preprint). The zone between the actual horizon and LSS may be called physical region, for we rely heavily on our physical theories in inferring its structure. Finally, beyond LSS and approaching the initial singularity (whatever its nature may be) one substitutes gradually positive science methodology and makes increasingly use of speculative approach. We call this zone "beyond horizon" metaphysical region, for obvious reason.

One might add to this picture two more characteristic distances (if not horizons), those of the beginning and end of the inflatory phase, but this would not make the matter more transparent at all. This phase is characterized, by definition, by chaotic "structure", endowed with uncontrollable fluctuations of the primordial fields. The very notion of information looses its meaning for this cosmic phase, as we have seen before. From this phase to the very Big Bang instance (the event horizon) we have in principle inaccessible region, what makes the theoretical efforts essentially speculative ones (metaphysical phase). Physical horizon (event horizon) is defined by $r_{ph} = T_a c$, where $T_a$ is estimated (according to the HBB model) age of the universe. No light that was emitted from beyond this distance could have reached us yet (absolute horizon). The same hold for other possible ways of energy transmission, like gravitational one, with proviso that they propagate with the same velocity (e.g. $c_{em} = c_g$). Hence, it is time that is limiting factor, not the space, more precisely, the finite age of our universe sets boundary to its spatial sector. This sort of limitation is absent in the Abderian paradigm, with an infinitely old universe. Of course, Olbers' paradox remains as an insurmountable obstacle in this picture, but not in Anaxagoras' hierarchical paradigm, as we have noted before.

## 12. OBSERVATIONAL EVIDENCE AND PHYSICAL REALITY

Empirical evidence of the surrounding physical environment comes to our mind via our senses. By far the most of this evidence we acquire by electromagnetic radiation (mainly by light), and we consider this sense as the most reliable, though not absolutely so. In everyday life we occasionally resort to other senses, tactile, for instance (as the case of "untrustful Thomas" illustrates).

Our empirical evidence relevant for constructing and verifying cosmological models comes from earthly data and astronomical observations (we are indebted to Referee for drawing our attention to this point). The former consist of various geological inferences, which set at least limits to cosmic parameters to deal with. It concerns, above all, the lower limit of the age of the present-day universe, which was a point of strong controversy in $19^{th}$ century. The present-day estimate of the age of our Earth has been the most effective tool in discarding various nonscientific claims concerning cosmogony, like those raised by so-called creationists (see, e.g., Brush, 1983, 2000). The claims by their descendents, Intelligent Designer proponents, are resistant to such arguments (as they are to any), but this is not a problem of cosmology (or any other serious discipline, for that matter).

In astronomy, apart from the massive cosmic rays, we must rely on our visual evidence almost exclusively. But we know both from everyday experience and laboratory experiments that the eyes can be deceiving. So-called optical illusions are cases in point, as skillfully illustrated on Escher's drawings, for instance.

Strictly speaking one may speak of two kinds of optical illusions. First, we may have correct observations, but wrong interpretations of the visual evidence. The case in point was the geocentric planetary (Ptolemaic) model, which reigned for some 16 centuries in Europe. The more elaborate and more scientific model, that of Aristarchos from Samos, was only partially accepted at the time, to be resurrected by Copernicus 18 centuries later. The crucial point was the assumption on Earth kinematics status as a celestial body, which in its turn depended on the dynamics available at the time. That the leading scientific minds, like Hipparchus and Archimedes had a correct picture of the latter has become evident now (e.g. Russo, 2002), but this did not prevent the subsequent generations, particular Middle-Age Europeans, to proclaim the Aristotelian picture a scientific dogma (so as to comply with the existing religious one), what stopped development of the cosmology for almost two millennia. Empirical evidences without theoretical model may be very deceiving, indeed.

We now turn to the optical illusions proper. In cosmology it translates into the question of the nature of the cosmic space topology (see. e.g. Luminet, 2001). As we have seen in the previous section, our empirical universe is less than the actual one, according to the Friedmann-Lemaitre paradigm. But one may envisage cosmos with a topology which provides the cosmos that is much smaller than the visual evidence suggests. We may see cosmic objects, like galaxies, clusters etc many times, like an object in a room with mirrors. This sort of multiple connected topology (so-called crumpled universe) has been contrived by a number of authors, like Weeks (see. e.g. Luminet *et al.* 1999) and it remains to find out which topology our universe has adopted before drawing conclusions from the visual evidence on the real structure (and extent) of the cosmos we live in.

In the binary-brane model due to Steinhardt (2002), mentioned before, an oscillatory universe has been contrived, with cyclic motions along both bulk sector and along the branes. The latter alternating expansions and contractions are experienced by the brane inhabitants, but for an outsider (sic), the brane expansion will never stop and the cosmos ends in an unlimited thinning of its content, approaching the state of an empty universe. Hence, the cyclic paradigm turns into an eternally expanding model, like the open Friedmann-Lemaitre one.[14]

Finally we mention a recent conjecture due to Ribeiro (2001) concerning the issue of verification of the assumed fractal structure of the visible universe. It suggests that if relevant observables are evaluated along the past light cone some cosmic characteristics assumed within the SM will appear the same as those ascribed to fractal structure. This possibility, apart from the practical significance in observational cosmology, illustrates well how much the observational methodology might be tightly bound with the data observed. What brings us back to the old question of the correlations between ontology and epistemology in general.

### 13. EPILOGUE

We have shown that many principal cosmological models, even paradigms, may be traced back to the Pre-Socratic Greece. That our knowledge and inferences into cosmic structure and evolution still relies on the ancient thinking should not be taken as a sign of stagnation in the field, but rather as the signature of fundamentality of the subject. In analyzing the current models we have not restricted ourselves to the most popular ones, in particular the Standard Model, but have tried to encompass as many as possible conceptually, if not theoretically, possible models that are on the market today, following Boltzmann's advice that no rationally designed concept should be discarded in advance.

We have drawn particular attention to the paradigm of hierarchical cosmos, in particular that as conceived by Anaxagoras. The rationale for that has been twofold. First, this concept appears gaining impetus as the new observational evidence is accumulating. Second, most recent research on the history of the concept, especially concerning thinker from Clazomenae, has been published in French, German, Russian and Italian and appears almost unknown in the English-speaking part of the cosmological community.

We have emphasized the epistemological aspects of the transition from the observational research of our cosmos, as a positive science (proper science) towards more and more speculative approach, as one moves to the remote parts of space and, eventually, time. It is shown that it is here that the epistemological issues become not only relevant, but crucial in giving credit to often wild speculations concerning the global physical world we live in.

We have put aside many relevant issues of more philosophical colour, trying to remain on the side of positive science. Also, the question of the relation of cosmology with religious thought has been only touched. (We refer interesting readers to the more specific literature on the subject; see, e.g. Grünbaum, 2000).

Is it to expect that the ultimate picture of the universe is attainable by a civilization like ours?[15] Or, put it this way, is the concept of evolution the ultimate truth we already possess? May one put forward a kind of antianthropic principle that requires that there can not be ultimate knowledge of the global structure and dynamic of the Universe. The choice, obviously, transcends the positive science and purely rational thinking. What does not make the issue less interesting, of course.

ACKNOWLEDGMENTS

I am indebted to Prof. Milan Ćirković for numerous references, fruitful discussions and critical reading of the manuscript. Many authors have kindly supplied me with their papers without which this article would not have been written. Helpful criticism and valuable suggestions by referees are thankfully acknowledged too The work has been supported by Ministry of Science and Environment Protection of Serbia.

NOTES

[1] Kant's *Mundus inteligibilis*, as opposed to *Mundus phaenomenon.*

[2] Since 1998, the evidence accumulates that the dynamics of the universe is dominated by repulsive "dark energy" comprising about 70% of the total energy of the universe (e.g. Perlmutter *et al* 1999, Riess *et al* 2001, Bennett *et al* 2003). It has been known since Einstein and de Sitter that such form of energy – of which Einstein's classical cosmological constant is a particular example – will cause a (quasi)exponential expansion after some specific epoch.

[3] As is well known the first to attack this law were Bohr *et al* (1924), while considering β decay, who badly failed in that respect. General relativity also disrespects energy conservation.

[4] We note that both Newtonian and GR equations have self-similar solutions, see, e.g. Harada and Maeda, 2001.

[5] The mathematical properties of the series 1:2:3 are not insignificant, either. Number 6, which can be decomposed into the sum and product of its divisors was valued particularly by Ancients, and was the basis of many number systems at the time. Philo of Alexandria argues in his treatise "On the Account of the World's Creation Given by Moses", in Greek *and Roman Philosophy after Aristotle*, J. Saunders (Ed.), The Free Press, New York, 1994, pp. 200-227, that God chose not by accident to create the World in 6 days, invoking the same mathematical properties of number 6 we mentioned.

[6] In fact, Anaxagoras' hierarchical model appears a generalization of this simple equivalence relation, applying it in an iterative manner, as we shall see later on.

[7] As Darwin's selection principle about survival of the fittest sounds to some epistemologists.
[8] It is for this reason Greeks refused to deal with it, even at the symbolic level (they had no sign for zero).
[9] Here and otherwise we are interested primarily in the epistemological aspects of the ancients' cosmologies. As for their actual or possible contents, which is still only partially known to us, see, e.g. Russo (2004)
[10] Only fragments of Anaxagoras' unique treatise *On Nature* have reached us. This is just a tentative reconstruction of his cosmology.
[11] Which has something to do with Eliade's usage of the same term (Eliade, 1964) but should not be identified with it.
[12] Following the pattern microcosm ~ macrocosm this Indian concept renders the belief in rebirth of animated beings.
[13] One could pursue the cosmology-mythology parallel even further, in particular concerning methodological analogy (see, e.g. Veyne, 1983), but this would lead us beyond the scope of the present article.
[14] In fact, this sort of time dilatation effect we have already had for the case of falling into black hole, finite in time for the astronaut, but infinitely slow for an external observer.
[15] See, e.g. the inspiring chap. 2 in Barrow, 1998.

Petar V. Grujić
*Institute of Physics*
*Pregrevica 118*
*11080 Zemun*
*Serbia*
*e-mail: grujic@phy.bg.ac.yu*